\documentclass[aps,prc,twocolumn,showkeys,showpacs]{revtex4}
\usepackage{graphicx,colordvi}
\usepackage{amssymb}
\usepackage{xcolor}
\begin{document}

\title{Mass Yields of Fission Fragment of Pt to Ra Isotopes}

\author{Krzysztof Pomorski$^1$, Artur Dobrowolski$^1$, Rui Han$^1$, 
Bo\.zena Nerlo-Pomorska$^1$,\\ Micha{\l} Warda$^1$, Zhigang Xiao$^2$, 
Yongjing Chen$^3$, Lilie Liu$^3$, Jun-Long Tian$^{4}$} 

\affiliation{
$^1$ Institute of Physics, Maria Curie Sk{\l}odowska University,
     20-031 Lublin, Poland\\
$^2$ Physics Department, Tsinghua University, Beijing 100084, China\\
$^3$ China Institute of Atomic Energy, Beijing 102413, China\\
$^4$ School of Physics and Electrical Engineering, Anyang Normal University, 
     Anyang 455000, China}

\pacs{21.10.Dr,25.70.Ji,25.85.-w,25.85.Ec}

\keywords{Nuclear fission, macroscopic-microscopic model, potential energy
          surfaces, fission fragment mass distributions}

\date{\today}

\begin{abstract}
An effective Fourier nuclear shape parametrization which describes well the most
relevant degrees of freedom on the way to fission is used to construct a 3D
collective model. The potential energy surface is evaluated within the 
macroscopic-microscopic approach based on the Lublin-Strasbourg Drop (LSD)
macroscopic energy and  Yukawa-folded single particle potential. A
phenomenological inertia parameter is used to describe the kinetic properties
of the fissioning system. The fission fragment mass yields are obtained by using
an approximate solution of the underlying Hamiltonian. The predicted mass
fragmentations for even-even Pt to Ra isotopes are compared with available
experimental data. Their main characteristics are well reproduced when the neck
rupture probability dependent on the neck radius is introduced.
\end{abstract}

\maketitle

\section{Introduction}

A proper reproduction of the fission fragments mass distribution (FMD) is one of
the most important tests of any theoretical model describing the nuclear fission
process. A very nice review of the existing fission models can be found in 
Ref.~\cite{MJV19}, which is dedicated to the memory of Arnie J. Sierk, one of
the leaders in this field of physics. Readers who are interested in the theory
of nuclear fission can find more details in the textbook \cite{KPo12}. So, we
are not going to recall similar information here. 

In the present paper we obtain such a distribution by an approximate solution of
the eigenproblem of the three-dimensional collective Hamiltonian, which
corresponds to the fission, neck, and mass-asymmetry modes, respectively. The
nonadiabatic and dissipative effects in low energy fission were taken into
account in a similar way as in Refs.~\cite{NPW89,NPI15,PIN17,PNB18}. The
potential energy surface (PES) is obtained by the macroscopic-microscopic
(mac-mic) method, where  the Lublin-Strasbourg Drop (LSD) model \cite{PDu03} has
been used for the macroscopic part of the energy, while the microscopic shell
and pairing corrections have been evaluated through the Yukawa-folded (YF)
single-particle levels \cite{DNi76,DPB16}. The shape of the fissioning nucleus
was described by the three-dimensional Fourier parametrization
\cite{PNB15,SPN17}. It was shown in Refs.~\cite{PNB15,Bar20} that this
parametrization described very well the shapes of the nuclei even those close to
the scission configuration.

The paper is organized in the following way. First we present shortly the
details of the theoretical model, then we show the collective potential energy
surface evaluated by the mac-mic model for the selected Pt to Ra isotopes. The
calculated fission FMD's are compared with the existing experimental data in the
next section. Conclusions and plans of further calculations are presented in
Summary.

\section{Model of the fission dynamics}

The evolution of a nucleus from the equilibrium state towards fission is
simulated by a simple dynamical approach based on the  PES, which depends on
three relevant collective degrees of freedom describing the
nuclear shape in this process: elongation of the nucleus ($q_2$), asymmetry of left and right
mass fragments ($q_3$), and the neck size ($q_4$). As demonstrated in
Refs.~\cite{PNB15,SPN17}, the shape parametrization of the deformed nucleus,
which gives an expansion of the nuclear surface in the form of the Fourier
series of dimensionless coordinate $(z-z_{\rm sh})/z_0$: 
\begin{equation}
\begin{array}{ll}\displaystyle 
\frac{\rho_s(z)^2}{R_0^2}=\sum\limits_{n=1}^{\infty} 
 & \bigg[ a_{2n} \cos\left(\frac{(2n-1)\pi}{2}\frac{z-z_{\rm sh}}{z_0}\right)\\
 & + a_{2n+1} \sin\left(\frac{ 2n\pi}{2}\frac{z-z_{\rm sh}}{z_0}\right) \bigg]
\end{array}
\label{rho2}
\end{equation}
\noindent 
is rapidly converging. As in the famous ``Funny-Hills'' parametrization 
\cite{BDJ72}, $\rho_s(z)$ defines the distance of the
surface point from the $Oz$ symmetry axis, and $z_0$ is the
half elongation of a nucleus between extreme points located at
$z_{\rm min} = z_{\rm sh} - z_0$ and
$z_{\mathrm{max}} = z_{\rm sh} + z_0$. The quantity
$z_{\rm sh}$ is responsible for shifting the center
of mass of axially symmetric nuclear drop to be located at the origin of
coordinate system. $R_0$ represents the radius of the
corresponding spherical nucleus of the same volume. In  Eq.
(\ref{rho2}), the parameters $a_2$, $a_3$, $a_4$ are related
to the $q_2$, $q_3$, $q_4$ deformation
parameters through the following formulas, respectively,
\begin{equation}
\begin{array}{ll}\displaystyle 
q_2=a_2^{0}/a_2-a_2/a_2^{0}, & q_3=a_3, \\[1ex] 
q_4=a_4+\sqrt{(q_2/9)^2+(a_4^{0})^2}, & q_5=a_5-(q_2-2)\frac{a_3}{10},\\[1ex]
q_6=a_6-\sqrt{(q_2/100)^2+(a_6^{0})^2}.&
\end{array}
\end{equation}
These relations proposed in Ref.~\cite{SPN17} transform the original deformation
parameters $a_i$ to the more natural parameters $q_i$, which ensure that only
minor variations of the liquid drop fission paths occur around $q_4=0$. In
addition, more and more elongated prolate shapes correspond to decreasing values
of $a_2$, while oblate ones are described by $a_2>1$, which is in contradiction
to the traditional definition of quadrupole deformation. Here $a_n^{(0)}$ stands
for the value of the $a_n$ coefficient for a sphere:
\begin{equation}
a^{(0)}_{2n}=(-1)^{n-1}\,32/[\pi\,(2n-1)]^3~.
\end{equation}

Having defined the shape parametrization of nuclear surface, one can
now calculate the PES in 3D collective space. The nuclear deformation energies
are determined in the mac-mic approach, where the smooth energy part is given by
the Lublin-Strasbourg Drop (LSD) model \cite{PDu03} and the microscopic effects
have been evaluated through a Yukawa-folded (YF) single-particle potential
\cite{DNi76, DPB16}. The Strutinsky shell-correction method
\cite{Str66,NTS69,BDJ72} with a $6^{\mathrm{th}}$ order correctional polynomial
and a smearing width $\gamma_S=1.2\hbar\omega_0$ is used, where
$\hbar\omega_0=41/A^{1/3}$ MeV is the distance between the spherical
harmonic-oscillator major shells. The BCS theory \cite{BCS57} with the
approximate GCM+GOA-like particle number projection method \cite{GPo86} is used
for the pairing correlations.

The pairing strengths $G{\mathcal N^{2/3}}=0.28\hbar\omega_0$, with ${\mathcal
N}=Z,N$ for protons or neutrons, was adjusted to the experimentally measured
mass differences of nuclei in this region with a ``pairing window'' containing
$2\sqrt{15{\mathcal N}}$ mean-field time-dependent degenerated levels lying
around the Fermi level \cite{PPS89}. The mean-field used to generate the
single-particle energy levels, entering the Strutinsky and BCS quantum
correction methods, is taken in the form of the folded Yukawa potential
\cite{DNi76} and diagonalized in the deformed harmonic-oscillator basis with 18
major shells as written in Ref.~\cite{DPB16}.

The present research is a continuation and extension of our previous works
\cite{PNB18,NPW89,NPI15,PIN17}. The fundamental idea of the fission dynamics
discussed in this work is that the relatively slow motion towards fission,
mainly in  $q_2$ direction, is accompanied by the fast vibrations in the
``perpendicular'' $q_3$ and $q_4$ collective variables. This allows us to treat
both of these two types of motion as decoupled which, in consequence, gives the
wave function corresponding to the total eigenenergy $E$ of fissioning nucleus
approximately as\\[-3ex]
\begin{equation}
 \Psi_{nE}(q_2,q_3,q_4)=u_{nE}(q_2)\,\phi_n(q_3,q_4;q_2)~.
\end{equation}
The function $u_{nE}(q_2)$ is the eigenfunction corresponding to the motion
towards fission which depends mainly on a single variable $q_2$, while the
$\phi_n(q_3,q_4;q_2)$ simulates the $n-$phonon ``fast'' collective vibrations on
the perpendicular 2D $\{q_3,q_4\}$ plane for a given elongation $q_2$.

There is a proper way to determine the $u_{nE}(q_2)$ and $\phi_n(q_3,q_4;q_2)$
wave function components in the discussed 3D collective space, respectively. For
$u_{nE}(q_2)$ one can use the WKB approximation for a single $q_2$ mode as it
has been done in Ref.~\cite{PIN17}, in which a 2D collective space has been 
considered, only. For $\phi_n(q_3,q_4;q_2)$, one can solve the eigenproblem of
the underlying Hamiltonian in the perpendicular directions numerically. However,
when limiting only to the low energy fission, the density of probability
$W(q_3,q_4;q_2)$ of finding the system for a given elongation $q_2$, within the
area of $(q_3\pm dq_3, q_4\pm dq_4)$, is given as 
\begin{equation}
W(q_3,q_4;q_2)= |\Psi(q_2,q_3,q_4)|^2=|\phi_0(q_3,q_4;q_2)|^2~.
\label{W_prob}
\end{equation}

A further simplification of the model is to approximate the modulus
square of the total wave function in Eq.~(\ref{W_prob}) by the Wigner function
in the form of
\begin{equation}
W(q_3,q_4;q_2)\propto \exp{\frac{V(q_3,q_4;q_2)-V_{\rm min}(q_2)}{E_0}}~,
\label{W_prob1}
\end{equation}
where $V_{\rm min}(q_2)$ is the minimum of the potential for a given elongation
$q_2$, and $E_0$ is the zero-point energy which is treated as an
adjustable parameter.

To obtain the fragment mass yield for a given elongation $q_2$ one has to
integrate the probabilities (\ref{W_prob1}) coming from different neck shapes,
simulated basically by the $q_4$ parameter\\[-3ex]
\begin{equation}
w(q_3;q_2)=\int W(q_3,q_4;q_2) dq_4~.
\label{W_integr}
\end{equation}
It is clear that the fission probability may strongly depend on the neck
thickness, strictly speaking, its radius $R_{\rm neck}$. Following the idea from
Ref.~\cite{PIN17} one assumes the neck rupture probability $P$ to be equal to
\begin{equation}
P(q_2,q_3,q_4)=\frac{k_0}{k}\,P_{\rm neck}(R_{\rm neck})~,
\label{neck_rap}
\end{equation}
where $P_{\rm neck}$ is a geometrical factor indicating the neck breaking
probability proportional to the neck thickness, while $k_0/k$ describes the fact
that the larger collective velocity towards fission, $v(q_2)=\dot q_2$, gives
the less probable neck rupture. The constant parameter $k_0$ plays the role of
scaling parameter which is finally eliminated when calculating the resulting
FMD. The expression for  the geometrical probability factor $P_{\rm neck}(R_{\rm
neck})$ can be chosen in an arbitrary way to some extent,  however after a
number of trials we have used the Gaussian form \cite{PNB18}
\begin{equation}
P_{\rm neck}(R_{\rm neck})=\exp{[-\log 2(R_{\rm neck}/d)^2]}~,
\label{P_neck}
\end{equation}

\noindent 
where $d$ is the ``half-width'' of the probability and is treated here as
another adjustable parameter. The momentum $k$ in Eq.~(\ref{neck_rap}) simulates
the dynamics of the fission process which, as usually, depends both on the local
collective kinetic energy $E-V(q_2)$ and the inertia towards the leading
variable $q_2$
\begin{equation}
\frac{\hbar^2 k^2}{2\bar{M}(q_2)}=E_{\rm kin}=E-Q-V(q_2) ~,
\label{energy_cons}
\end{equation}
with $\bar{M}(q_2)$ standing for the averaged inertia parameter over $q_3$ and 
$q_4$ degrees of freedom at a given elongation $q_2$, and $V(q_2)$ is the
averaged potential. In the further calculations we assume that the part of the
total energy converted into heat $Q$ is negligibly small due to the very low
friction force when the collective velocity $v(q_2)$ is small in low energy
fission. A good approximation of the inertia $\bar{M}(q_2)$ is to use the
irrotational flow mass parameter $B_{\rm irr}$ \cite{RLM76}, which is derived
initially as a function of the single collective parameter $R_{12}$, the
distance between fragments, and the reduced mass $\mu$ of both fragments
\begin{equation}
\bar{M}(q_2)=\mu [1+11.5\,(B_{\rm irr}/\mu -1)]\bigg(\frac{\partial R_{12}}
{\partial q_2}\bigg)^2 ~.
\label{irrM}
\end{equation}

In order to make use of the neck rupture probability $P(q_3,q_4;q_2)$ in
Eq.~(\ref{neck_rap}), one has to rewrite the integral over $q_4$ probability
distribution (\ref{W_integr}) in the form of
\begin{equation}
w(q_3;q_2)=\int W(q_3,q_4;q_2) P(q_2,q_3,q_4) dq_4 ~,
\label{W_integrP}
\end{equation}
in which now the neck rupture probability is, in addition, taken into account.
The above approximation implies a very important fact that, for a fixed $q_3$
value, the fission may occur within a certain range of $q_2$ deformations with
different probabilities. Therefore, to obtain the true fission probability
distribution $w'(q_3;q_2)$ at a strictly given $q_2$, one has to exclude the
fission events occurred in the ``previous'' $q'_{2}<q_2$ configurations, i.e.,
\begin{equation}
w'(q_3;q_2)=w(q_3;q_2)\frac{1-\int\limits_{q'_{2}<q_2} w(q_3;q'_2) dq'_2}
{\int w(q_3;q'_2) dq'_2}~.
\label{w_prev}
\end{equation}

The normalized mass yield is then obtained as the sum of partial yields at
different given $q_2$:
\begin{equation}
Y(q_3)=\frac{\int w'(q_3;q_2) dq_2}{\int w'(q_3;q_2) dq_3 \, dq_2}~.
\label{Yield}
\end{equation}

Since there is an one-to-one correspondence between $q_3$ deformation and the
masses of the left ($A_L$) and right ($A_R=A-A_L$) fission fragments, the yield
function of Eq.~(\ref{Yield}) can be directly compared with the experimental
FMD's now. One should notice that the scaling parameter $k_0$ introduced in Eq.
(\ref{neck_rap}) does not longer appear in the definition of mass yield. Therefore,
the only free parameters of the above model are: zero-point energy parameter
$E_0$ in Eq.~(\ref{W_prob}) and the half-width parameter $d$ appearing in the
probability of neck rupture (\ref{P_neck}). 
\begin{figure}[h!]
\includegraphics[width=0.9\columnwidth]{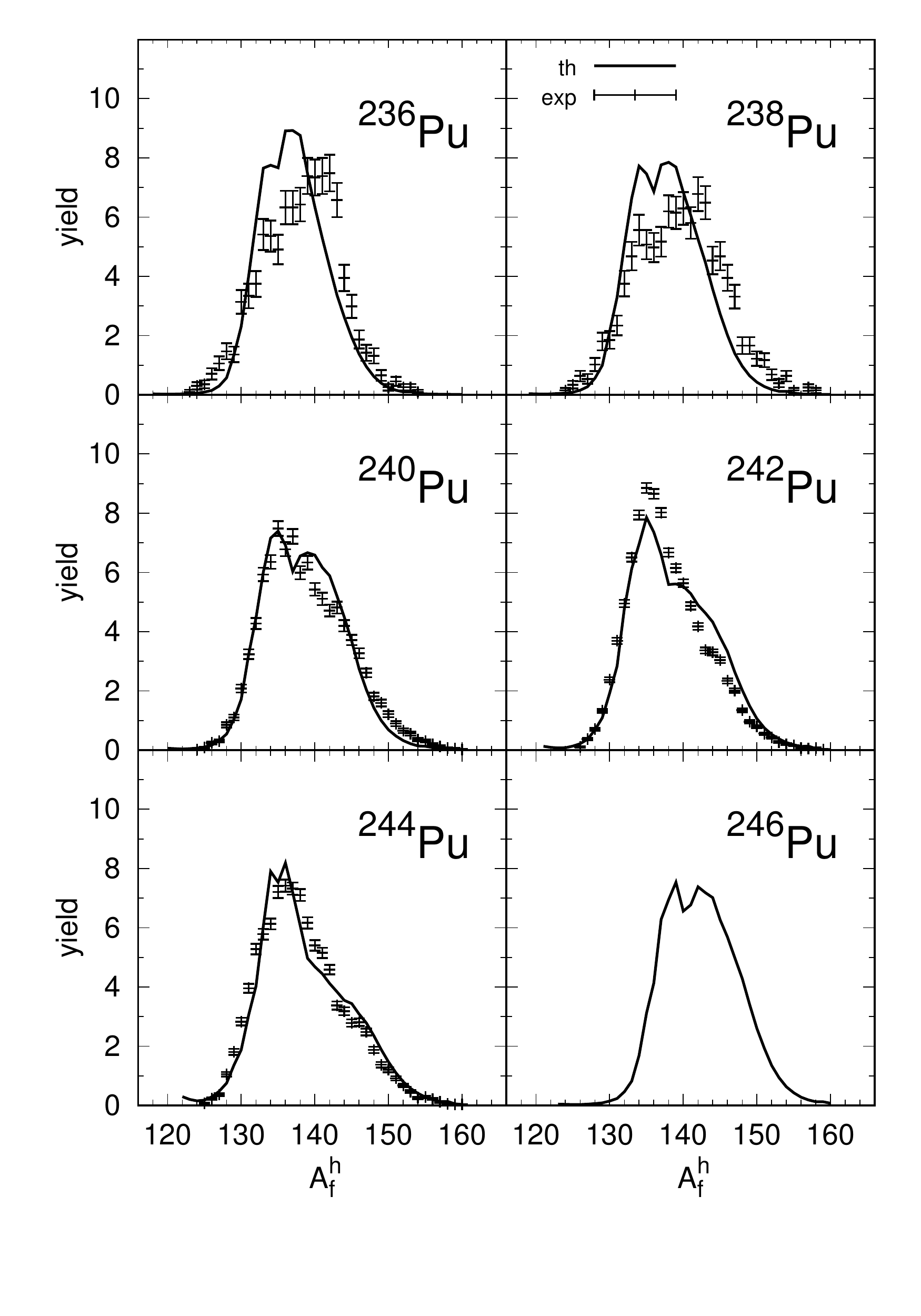}
\vspace{-1.5cm}

\caption{Theoretical estimates (solid lines) of heavy fragment mass yield 
for fission of $^{236-246}$Pu isotopes compared with the experimental data
(points with bars) from Ref.~\cite{Dem97}. }
\label{Pu}
\end{figure}

In Ref.~\cite{PNB18}, the parameter $d=0.15 R_0$ was adjusted to reproduce the 
experimental fragment mass yields measured in the low energy fission of 
$^{236-244}$Pu isotopes. On the other hand, zero-point energy parameter $E_0 =
1$ MeV related  to the $q_3$ and $q_4$ degrees of freedom is kept constant. The
comparison of the estimates obtained by the above model with the data taken from
Ref.~\cite{Dem97} is shown in Fig.~\ref{Pu}. It can be seen that the  agreement
of this model predictions with the experimental yields is  pretty good. 
\begin{figure}[h!]
\includegraphics[width=0.95\columnwidth]{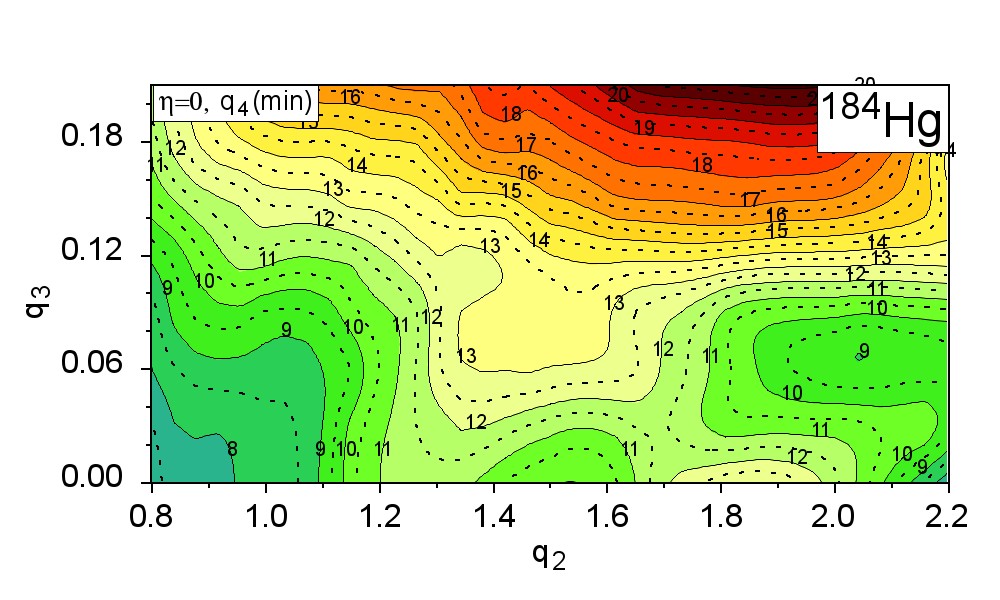}\\[-2ex]
\caption{Potential energy surface of $^{184}$Hg at the $(q_2,q_3)$ plane
minimized with respect to $q_4$.}
\label{184Hg23}
\end{figure}

\section{Results}

\begin{figure}[t!]
\includegraphics[width=0.95\columnwidth]{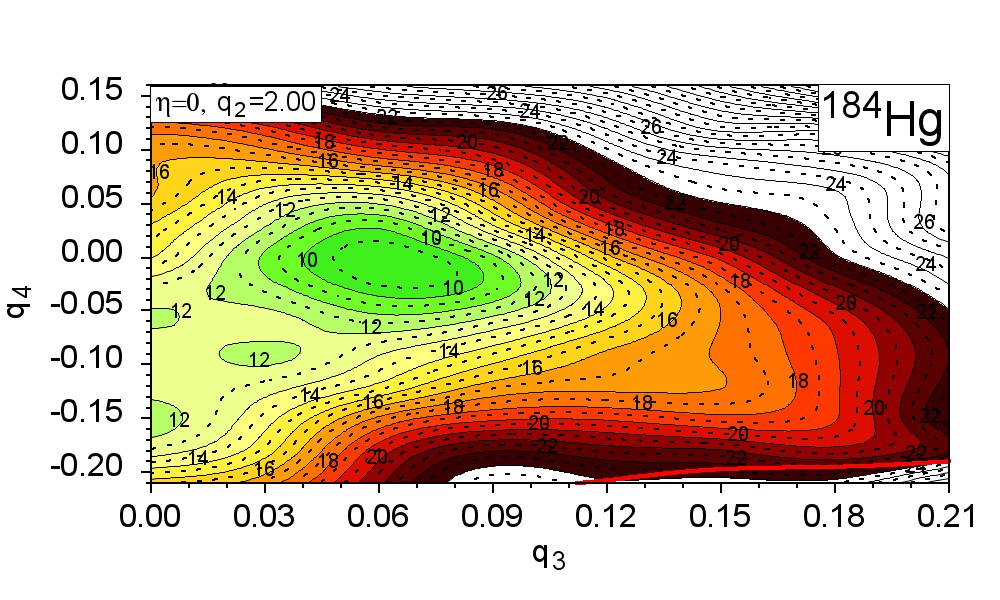}\\[-2ex]
\includegraphics[width=0.95\columnwidth]{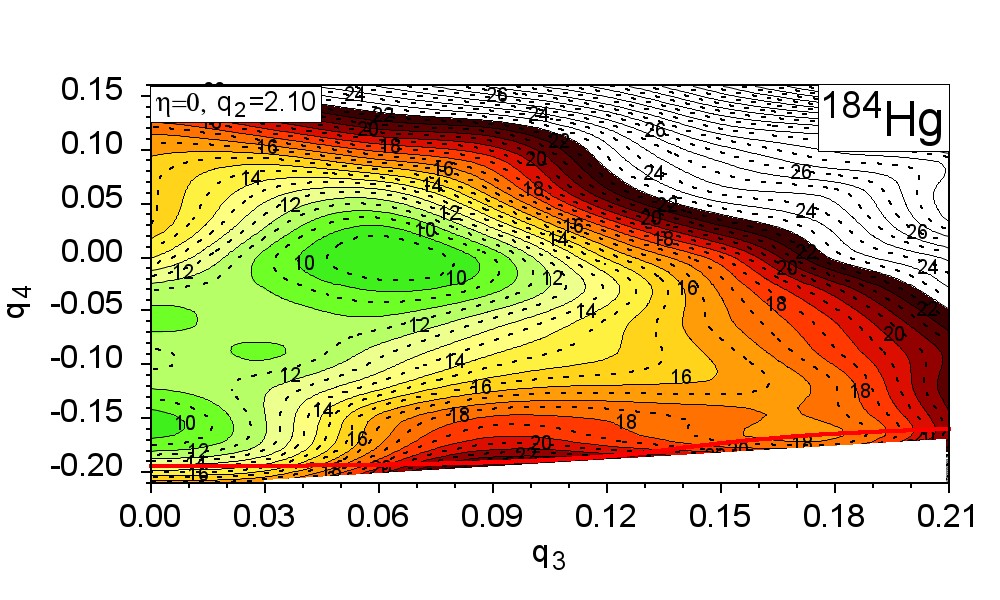}\\[-2ex]
\includegraphics[width=0.95\columnwidth]{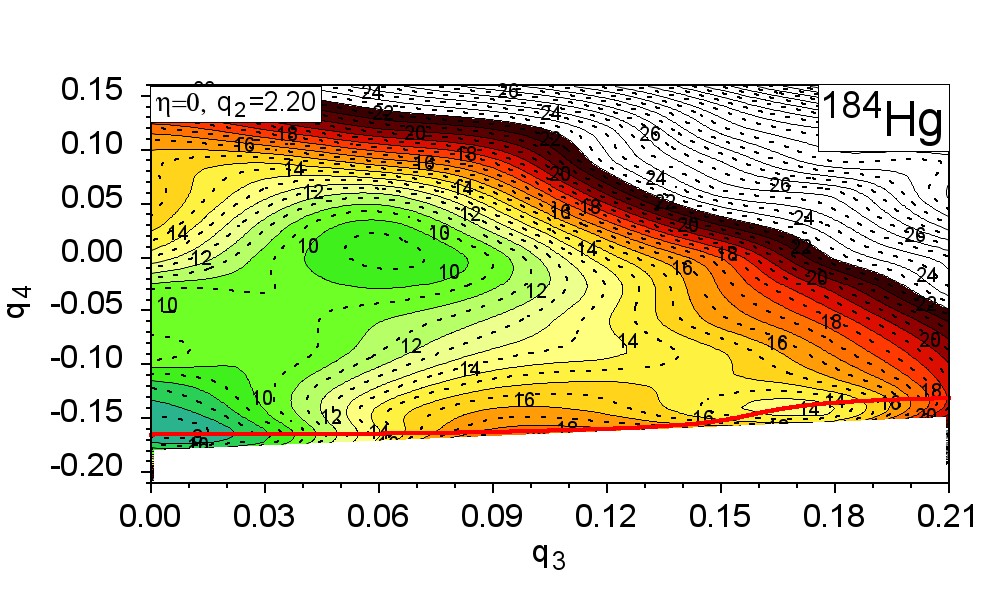}\\[-2ex]
\includegraphics[width=0.95\columnwidth]{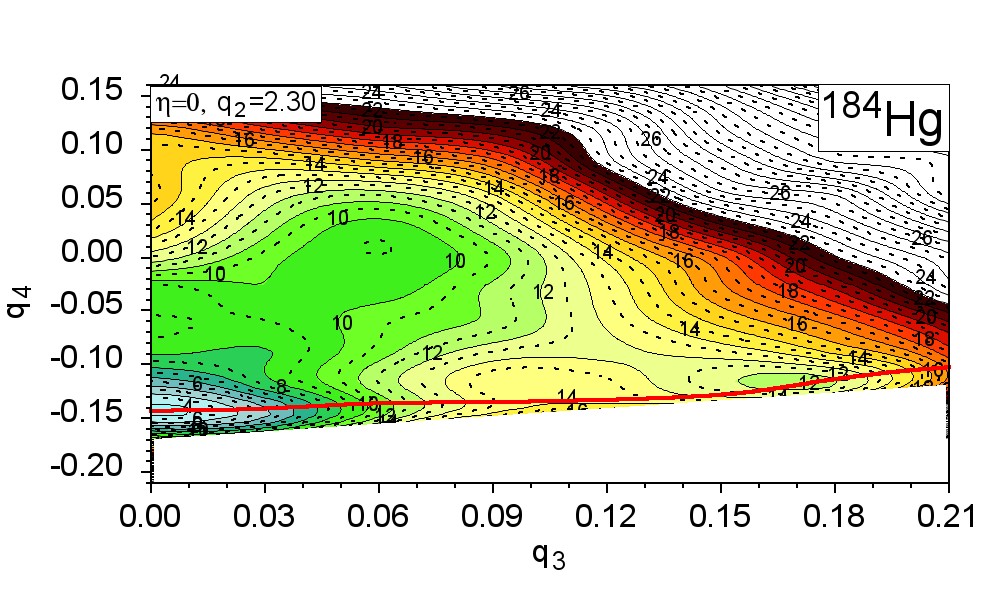}\\[-2ex]
\caption{Potential energy surface cross-sections of $^{184}$Hg on the plane
$(q_3,q_4)$. The panels from top to bottom correspond to elongations $q_2=$ 2.0,
2.1, 2.2, and 2.3, respectively. The solid red lines drawn in each pannel
correspond to the neck radius equaling to the nucleon radius.}
\label{184Hg34}
\end{figure}

The results obtained in Ref.~\cite{PNB18} for Pu isotopes have encouraged us to
investigate the possibilities of extending our model to predict/describe the
FMD's of the low energy fission of  Hg, Pt, Pb, Po, Rn and Ra nuclei. The
fission barriers of these nuclei around $^{208}$Pb are significantly higher than
those for actinides. In these nuclei a typical potential energy difference
between the energies of the saddle and the most probable scission point is of
the order of a few MeV, which is about only one order of magnitude smaller than
that in actinides. It is then clear that in these two nuclear regions, the role
of the fission dynamics should be significantly different.

The PES of all the considered nuclei are evaluated on the following 4D grid in
the deformation parameter space (2):
\begin{equation}
\begin{array}{lr}\displaystyle 
q_2=&-0.60~(0.05)~2.35 ~,\\
q_3=&0.00~(0.03)~0.21 ~,\\
q_4=&-0.21~(0.03)~0.21 ~,\\
\eta~=&0.00~(0.03)~0.21 ~,
\end{array}
\label{grid}
\end{equation}
where the parameter $\eta$  describes the non-axial shapes of nuclei as defined
in Ref.~\cite{SPN17}. It turns out that for all the considered isotopes
$\eta\neq 0$ may only appear in rather less elongated nuclei. Its influence
practically ends around the deformations corresponding to the first saddle point
in the fission barrier, i.e., at  $q_2\approx 1.3$. Since in the following we
will only consider the fission fragment mass distribution of non-rotating
nuclei, this non-axial mode will be neglected.
\begin{figure}[h!]
\includegraphics[width=0.95\columnwidth]{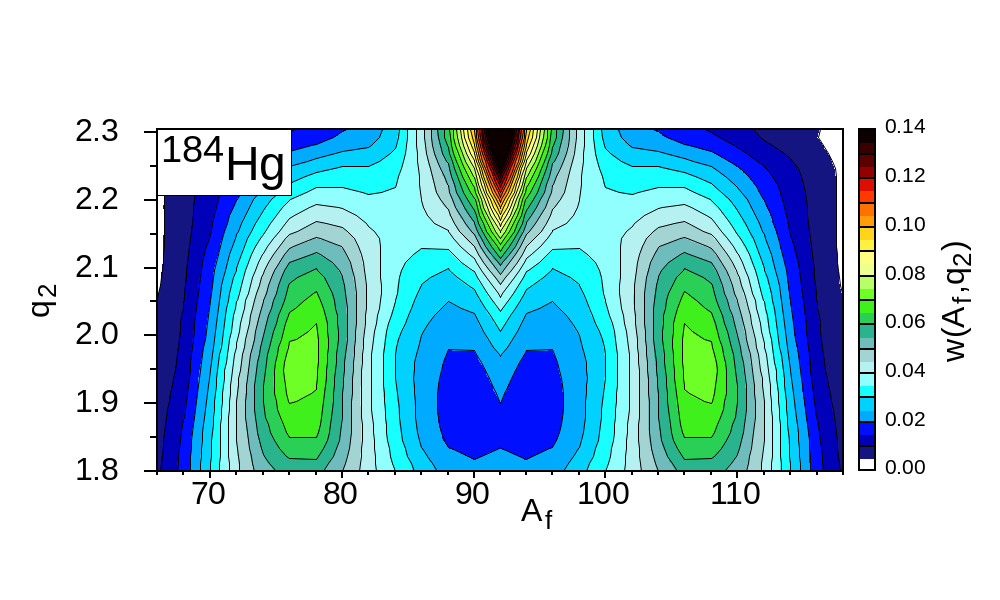}\\[-2ex]
\caption{Fission fragment distribution probability of $^{184}$Hg integrated over
 the deformation parameter $q_4$ (see Eq.~\ref{W_integr}) at the $(A_f,q_2)$
 plane.}
\label{184Hgw}
\end{figure}

A typical example of PES on the $(q_2,q_3)$ plane for $^{184}$Hg, where the 
macroscopic-microscopic energy of $^{184}$Hg is minimized with respect to the
neck degree of freedom  $q_4$, is shown in Fig.~\ref{184Hg23}. The labels at the
layers correspond to the energy (in MeV) with respect to the LSD macroscopic
energy of  spherical nucleus. The first saddle is noticed at $q_2=1.28$ and
$q_3=0$, while the second one is at  $q_2=1.69$ and $q_3=0.03$. Let us notice
that this 2D energy map is only a projection of the full 3D PES onto ($q_2,
q_3$) plane. A more complete PES structure of $^{184}$Hg can be observed in
Fig.~\ref{184Hg34}, where the ($q_3, q_4$) cross-sections corresponding to
different elongations $q_2$=2.0,  2.1, 2.2, and 2.3 are shown, respectively.

\begin{figure*}[htb]
\includegraphics[width=\textwidth]{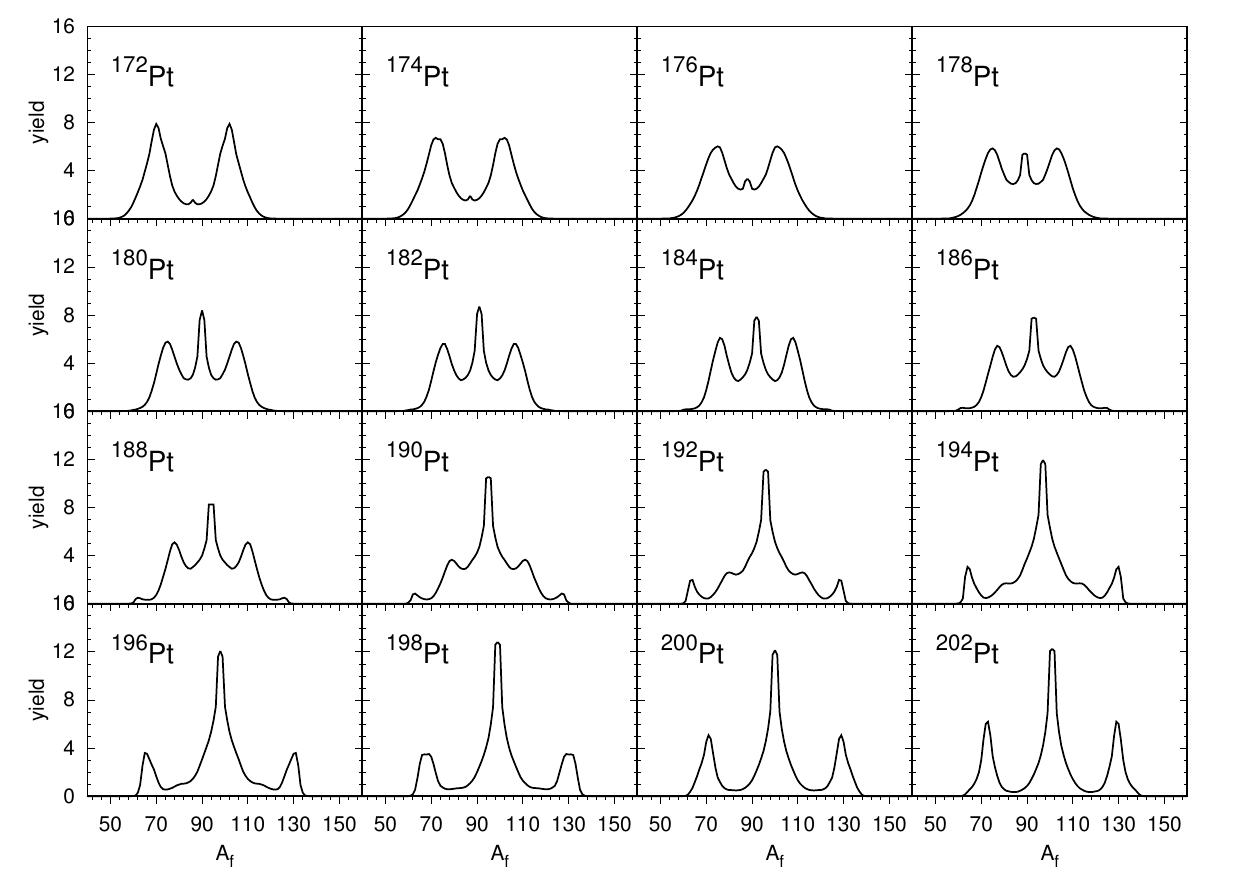}
\includegraphics[width=\textwidth]{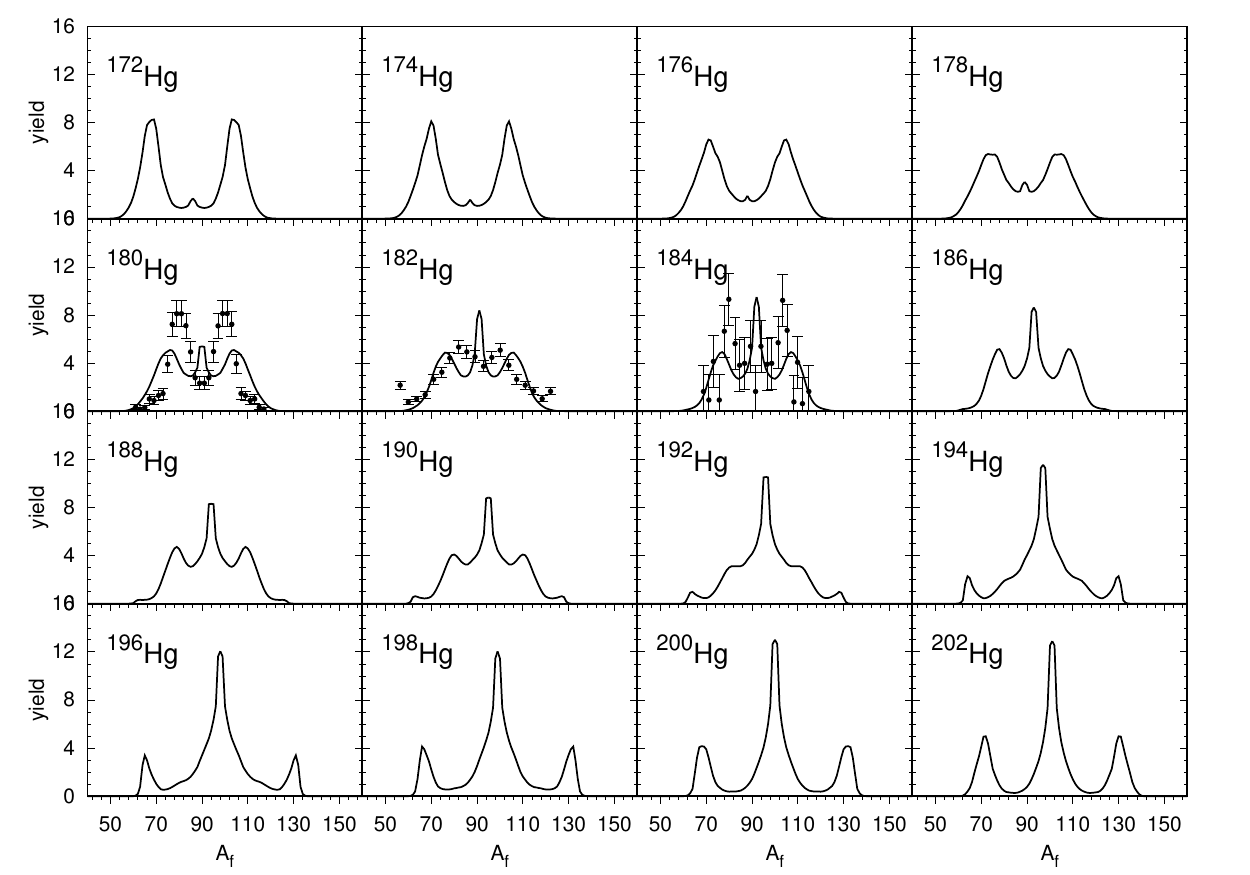}\\[-2ex]
\caption{Fission fragment mass yields of Pt (top part) and Hg (bottom part)
         isotopes. Experimental data (points with bars) are taken from 
         Refs.~\cite{Els13,Pra15,Gor13}.}
\label{Pt-Hg}
\end{figure*}
\begin{figure*}[htb]
\includegraphics[width=\textwidth]{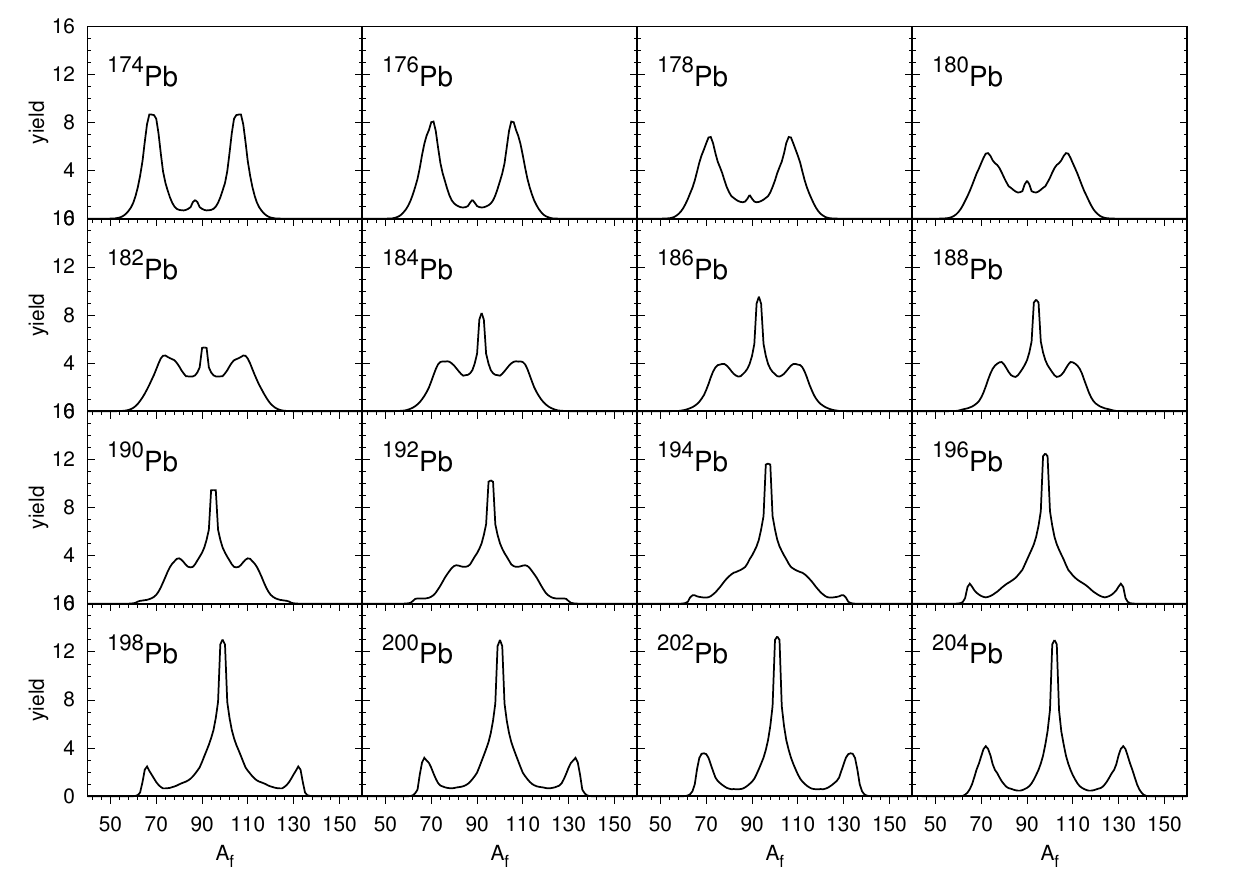}
\includegraphics[width=\textwidth]{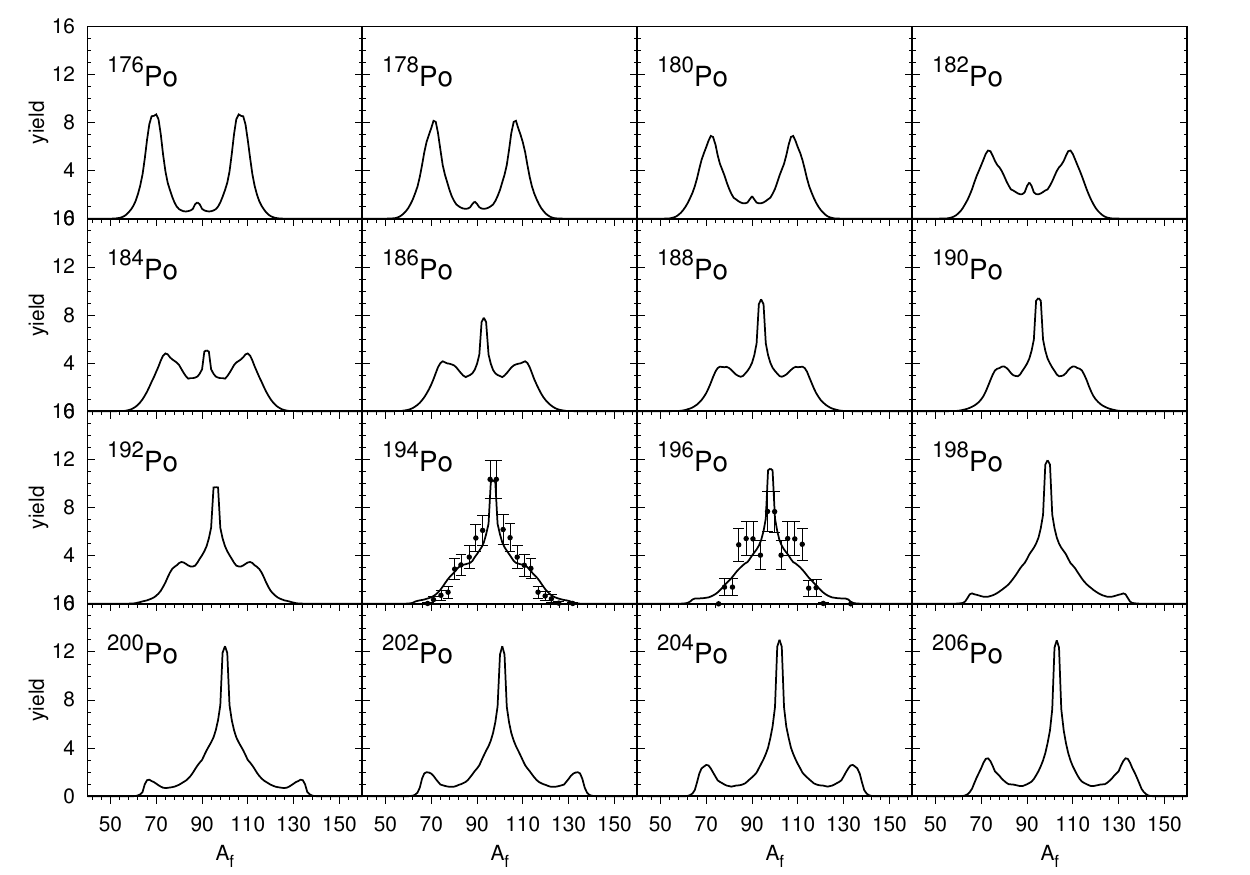}\\[-2ex]
\caption{Fission fragment mass yields of Pb (top part) and Po (bottom part)
         isotopes. Experimental data (points with bars) are taken from 
         Ref.~\cite{Ghy14}.}
\label{Pb-Po}
\end{figure*}
\begin{figure*}[htb]
\includegraphics[width=\textwidth]{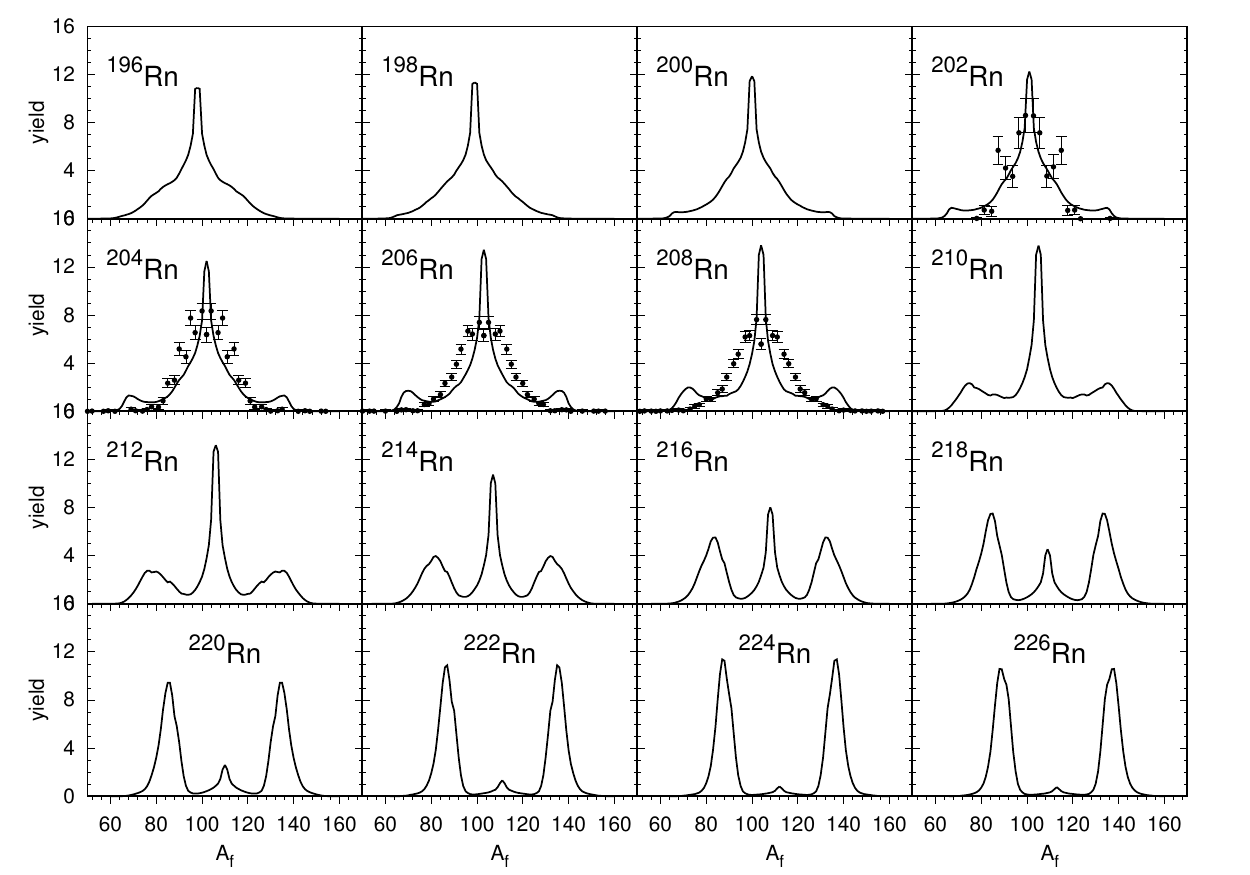}
\includegraphics[width=\textwidth]{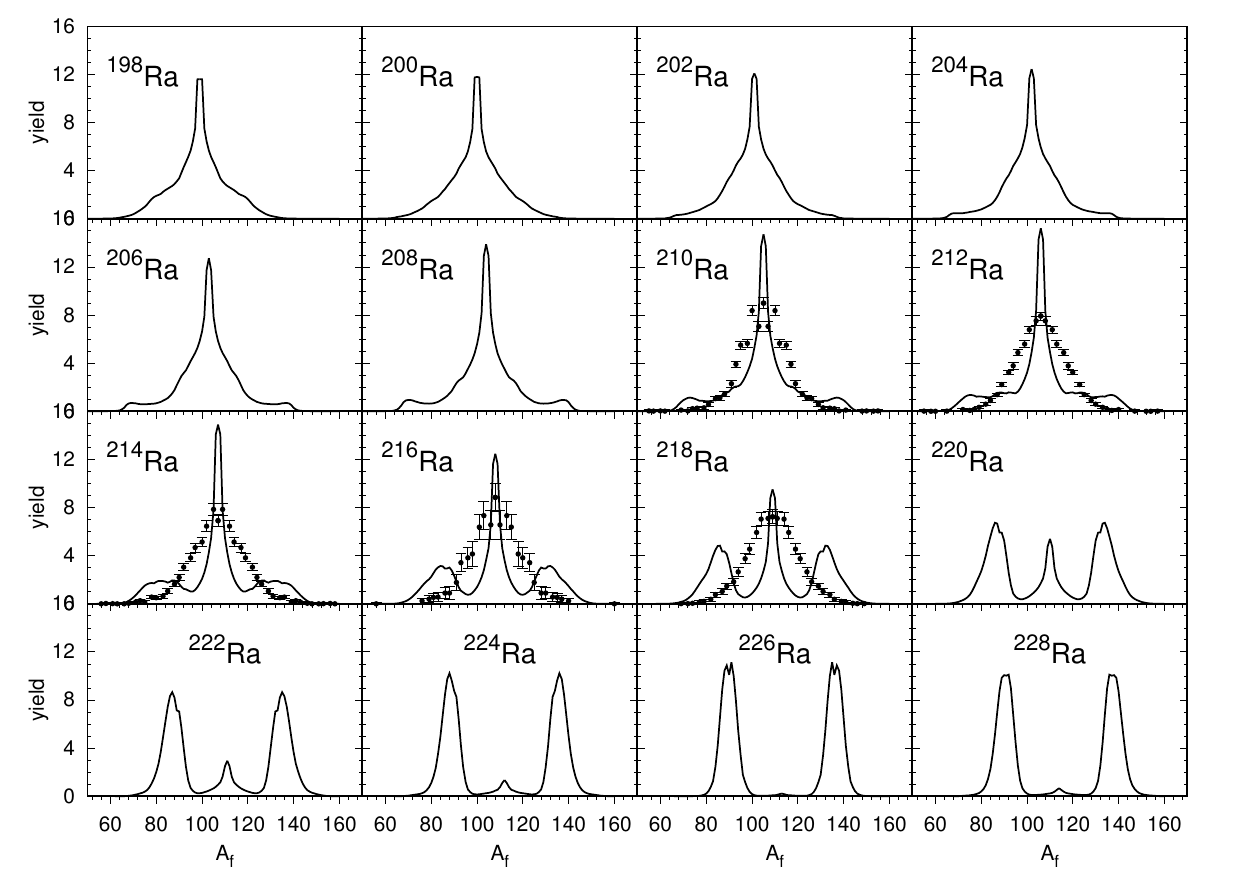}\\[-2ex]
\caption{Fission fragment mass yields of Rn (top part) and Ra (bottom part)    
         isotopes. Experimental data (points with bars) are taken from  
         Refs.~\cite{Ghy14,RMo13,Sch00}.}
\label{Rn-Ra}
\end{figure*}
\clearpage

\noindent
It can be seen that in Fig.~\ref{184Hg34} there are two competing minima, or
better to say, fission valleys: one corresponding to the asymmetric fission
around $q_3=0.06$ and $q_4=0$, and the second one towards symmetric  fission
around $q_3=0$ and $q_4=-0.15$. One can also see that with increasing $q_2$ the
symmetric valley becomes deeper. The solid red lines marked in maps of
Fig.~\ref{184Hg34} correspond to the liquid drop neck radius equaling to the
nucleon radius, which roughly approximate the scission lines.

The probability distribution (\ref{W_integr}) for different mass numbers ($A_f$)
as a function of the elongation $q_2$ is shown in Fig.~\ref{184Hgw}. It is seen
that at deformations $q_2\leq 2.15$, the asymmetric  fission with the heavier
fragment mass number $A_f\approx 106$ is the most probable mode, while at larger
elongations, the symmetric fission channel begins to dominate. The interplay
between these different fission modes depends on the fission dynamics and the
neck break probability as described in the previous section. Similarly as in 
Ref.~\cite{PNB18}, the $d$ parameter in Eq.~(\ref{P_neck}) is adjusted to the
known experimental FMD's \cite{Ghy14,Sch00}, while $E_0=$1 MeV (\ref{W_prob1})
remains  unchanged. The fitted half-width $d=1.5\,\,$fm according to the data of
Hg-Ra isotopes slightly differs from those adjusted to Pu nuclei in
Ref.~\cite{PNB18} ($d\approx 1.1\,\,$fm). This difference may be related to
different geometrical  features of fission barriers in these two regions. One
can say that in the considered Pt-Ra isotopes, the parameter $d$ tunes the
interplay between the asymmetric and  symmetric peaks in the FMD's. 

In Fig.~\ref{Pt-Hg} we can see the FMD's of Pt-Hg isotopes. In the examined
isotopic chains of even-even Pt nuclei, the FMD's gradually evolve from two-peak
asymmetric division with a non-zero admixture of symmetric fission towards the
division with dominating symmetric channel in the neutron deficient isotopes.
Moreover, the asymmetric channel strongly competes with the symmetric one in
$^{178-188}$Pt, while the latter one is gradually suppressed in $^{190-198}$Pt.

Similar behaviors exhibit FMD in Hg (Fig.~\ref{Pt-Hg}), Pb and Po 
(Fig.~\ref{Pb-Po})
isotopes. For the three experimentally measured isotopes $^{180-184}$Hg
\cite{Els13,Pra15,Gor13}, the theoretical FMD's asymmetric peaks are
underestimated distinctly, however, their overall shapes are reproduced. Note
that in contrary to the estimates made in Ref.~\cite{AAA13}, our model predicts 
the asymmetric fission to be the most probable mode in the lightest Hg isotopes.
Our prediction for $^{180}$Hg is in line with the experimental data \cite{Els13}
and in addition is confirmed by Ref.~\cite{Lib13}, where the asymmetric fission
mass distribution was deduced from total kinetic energy yield of the fragments.
The asymmetric peakes in the FMD for $^{182}$Hg is less visible, because the
measurement reported in Ref.~\cite{Pra15} was preformed at excitation energy
$E^*$=33.5 MeV, which slightly suppressed the shell effect.  A substantially better
reproduction of measured mass divisions is observed in Po isotopes. For
$^{194}$Po and $^{196}$Po, theoretical curves fit within the error-bars
of the experimental distributions taken from Ref.~\cite{Ghy14}.

In Fig.~\ref{Rn-Ra}, the predicted evolution of FMD's in Rn  isotopic chains is
completely opposite to the previously discussed Pt-Po even-even chains. Neutron
deficient $^{196-204}$Rn preferentially fission into two symmetric fragments,
while in $^{206-212}$Rn an asymmetric component is getting more and more
pronounced. The asymmetric peak becomes comparably high in moderately neutron
excessed $^{214-218}$Rn, while the symmetric peak in strongly neutron rich
$^{222-226}$Rn is suppressed significantly. In the experimentally measured
$^{202-208}$Rn isotopes, the dominating symmetric divisions are reproduced with
a tendency to slightly underestimate the asymmetric fission. Similar situations
appear in Ra isotopes. The rather broad symmetric FMD's, which are found 
experimentally
in $^{210-218}$Ra, is confronted with the rather narrow symmetric peaks and the
smaller asymmetric bump around $A_f=126$, which are predicted by our model. The
dominating asymmetric fission with heavier mass fragment around $A_f=138$ is 
predicted to be heaviest for the investigated Ra isotopes. 
In the cases when the experimental FMD's are available as functions of the
fragment charge ($Z_f$), we have simply assumed that $Z/A=Z_f/A_f$, where $Z$
and $A$ are the charge and mass numbers of the mother nucleus, respectively.

It must be pointed out that the experimental FMD's in the above Rn and Ra
isotopes correspond to their fission at $E^*\approx$18 MeV, i.e., about 10 MeV
above the saddle point, which corresponds to the initial temperature of
fissioning nucleus around $T=0.7$ MeV. At such temperature the pairing
correlations in nuclei become weaker or even disappear, which could influence
the width of symmetric and asymmetric valleys in the PES's. In addition, one has
to remind that there is no dissipation in our model and it is known that this
effect enlarges the width of the fission fragment yields.

\section{Summary and conclusions}

Summarizing our investigations we can write:
\begin{itemize}
\item a three-dimensional set of the Fourier deformation 
      parameters is sufficient to describe the properties of the fission
      process,
\item potential energy surfaces of nuclei are evaluated in the 
      macroscopic-microscopic model, where the LSD energy has been used 
      for the macroscopic smooth part, while the shell and pairing corrections
      are estimated on the basis of the Yukawa-folded single particle potential,
\item a collective 3D model based on the elongation, mass asymmetry and neck  
      modes is introduced,
\item a Wigner function is used to approximate the probability distribution
      related to the neck and mass asymmetry degrees of freedom,
\item a neck-breaking probability depending on the neck-size is introduced
      in order to reproduce the measured fission fragment mass yields.
\end{itemize}

It is shown that our collective 3D model which couples fission, neck and mass 
asymmetry collective modes is able to describe the main features of the fragment
mass yields in Pt-Ra and Pu isotopes. Due to its simplicity, this model may
serve for the rapid and pilot-type calculations of fission properties. To obtain
more precise results one has to use more advanced models in which the energy
dissipation and particle evaporation are taken into account in the fission
dynamics, e.g. the Langevin dynamics (confer Ref.~\cite{KPo12}) or the improved 
quantum molecular dynamics model (ImQMD), which has been successfully applied to 
describe the fission process in the heavy ion induced fission reactions, where
the excitation energy increases leading possibly to shorter fission time scale and 
even to the occurrence of a ternary fission \cite{QHW19,TWL10,LTQ13}.

Such calculations, based on the Fourier shape parametrization, as well
as on the self-consistent method, are planned to be carried out by our group in
the close future. 
\vspace{1cm}

\noindent
{\bf Acknowledgments}

This work is supported by the Polish National Science Center (Grant No.
2018/30/Q/ST2/00185) and by the National Natural Science Foundation of China
(Grant No. 11961131010), and the NSFC (Grants No. 11790325, 11875174 and
11890712). The authors would like to thank Christelle Schmitt for supplying us
with a part of the experimental data.

\end{document}